\documentclass[12pt]{article}
  \usepackage{amsfonts}
  \usepackage{amsmath}
\usepackage{amssymb}
\usepackage{amscd}
\usepackage{graphicx}
\begin{document}

~~
\bigskip
\bigskip
\begin{center}
{\Large {\bf{{{The Zeeman effect for hydrogen atom in twist-deformed space-time}}}}}
\end{center}
\bigskip
\bigskip
\bigskip
\begin{center}
{{\large ${\rm {Marcin\;Daszkiewicz}}$}}
\end{center}
\bigskip
\begin{center}
\bigskip

{ ${\rm{Institute\; of\; Theoretical\; Physics}}$}

{ ${\rm{ University\; of\; Wroclaw\; pl.\; Maxa\; Borna\; 9,\;
50-206\; Wroclaw,\; Poland}}$}

{ ${\rm{ e-mail:\; marcin@ift.uni.wroc.pl}}$}

\end{center}
\bigskip
\bigskip
\bigskip
\bigskip
\bigskip
\bigskip
\bigskip
\bigskip
\bigskip
\begin{abstract}
In this article we find the Zeeman corrections for hydrogen atom in the case of twist-deformed space-time. Particularly, we derive the corresponding orbital and spin $\hat{g}$-factors as well as we
notice, that the second one of them remains undeformed.
\end{abstract}
\bigskip
\bigskip
\bigskip
\bigskip
\eject

\section{Introduction}

The suggestion to use noncommutative coordinates goes back to
Heisenberg and was firstly  formalized by Snyder in \cite{snyder}.
Recently, there were also found formal  arguments based mainly  on
Quantum Gravity \cite{2}, \cite{2a} and String Theory models
\cite{recent}, \cite{string1}, indicating that space-time at the Planck
scale  should be noncommutative, i.e., it should  have a quantum
nature. Consequently, there appeared a lot of papers dealing with
noncommutative classical and quantum  mechanics (see e.g.
\cite{mech}, \cite{qm}) as well as with field theoretical models
(see e.g. \cite{prefield}, \cite{field}), in which  the quantum
space-time is employed.

In accordance with the Hopf-algebraic classification of all
deformations of relativistic \cite{class1} and nonrelativistic
\cite{class2} symmetries, one can distinguish three basic types
of space-time noncommutativity (see also \cite{nnh} for details):\\
\\
{ \bf 1)} Canonical ($\theta^{\mu\nu}$-deformed) type of quantum space \cite{oeckl}-\cite{dasz1}
\begin{equation}
[\;{ x}_{\mu},{ x}_{\nu}\;] = i\theta_{\mu\nu}\;, \label{noncomm}
\end{equation}
\\
{ \bf 2)} Lie-algebraic modification of classical space-time \cite{dasz1}-\cite{lie1}
\begin{equation}
[\;{ x}_{\mu},{ x}_{\nu}\;] = i\theta_{\mu\nu}^{\rho}{ x}_{\rho}\;,
\label{noncomm1}
\end{equation}
and\\
\\
{ \bf 3)} Quadratic deformation of Minkowski and Galilei  spaces \cite{dasz1}, \cite{lie1}-\cite{paolo}
\begin{equation}
[\;{ x}_{\mu},{ x}_{\nu}\;] = i\theta_{\mu\nu}^{\rho\tau}{
x}_{\rho}{ x}_{\tau}\;, \label{noncomm2}
\end{equation}
with coefficients $\theta_{\mu\nu}$, $\theta_{\mu\nu}^{\rho}$ and  $\theta_{\mu\nu}^{\rho\tau}$ being constants.\\
\\
Moreover, it has been demonstrated in \cite{nnh}, that in the case of the
so-called N-enlarged Newton-Hooke Hopf algebras
$\,{\mathcal U}^{(N)}_0({ NH}_{\pm})$ the twist deformation
provides the new  space-time noncommutativity of the
form\footnote{$x_0 = ct$.},\footnote{ The discussed space-times have been  defined as the quantum
representation spaces, so-called Hopf modules (see e.g. \cite{oeckl}, \cite{chi}), for the quantum N-enlarged
Newton-Hooke Hopf algebras.}
\begin{equation}
{ \bf 4)}\;\;\;\;\;\;\;\;\;[\;t,{ x}_{i}\;] = 0\;\;\;,\;\;\; [\;{ x}_{i},{ x}_{j}\;] = 
if_{\pm}\left(\frac{t}{\tau}\right)\theta_{ij}(x)
\;, \label{nhspace}
\end{equation}
with time-dependent  functions
$$f_+\left(\frac{t}{\tau}\right) =
f\left(\sinh\left(\frac{t}{\tau}\right),\cosh\left(\frac{t}{\tau}\right)\right)\;\;\;,\;\;\;
f_-\left(\frac{t}{\tau}\right) =
f\left(\sin\left(\frac{t}{\tau}\right),\cos\left(\frac{t}{\tau}\right)\right)\;,$$
$\theta_{ij}(x) \sim \theta_{ij} = {\rm const}$ or
$\theta_{ij}(x) \sim \theta_{ij}^{k}x_k$ and  $\tau$ denoting the time scale parameter
 -  the cosmological constant. Besides, it should be  noted, that the  above mentioned quantum spaces {\bf 1)}, { \bf 2)} and { \bf 3)}
can be obtained  by the proper contraction limit  of the commutation relations { \bf 4)}\footnote{Such a result indicates that the twisted N-enlarged Newton-Hooke Hopf algebra plays a role of the most general type of quantum group deformation at nonrelativistic level.}.

In this article we investigate the impact of the twisted N-enlarged
Newton-Hooke space-time \cite{nnh}\footnote{In the formula (\ref{relations}) $\kappa_a$
denotes the deformation parameter such that $\lim_{\kappa_a \to 0}f_{\kappa_a}(t)=0$.}
\begin{eqnarray}
[\;\hat{ x}_{1},\hat{ x}_{2}\;] = if_{\kappa_a}({t})\;,\label{relations}
\end{eqnarray}
on the spectrum of hydrogen atom in a weak magnetic field, with the electron spin taken into account \cite{kiku}. Particularly, we write the
energy operator describing the anomalous Zeeman effect for the nonrelativistic quantum space (\ref{relations}) and further, assuming that
the function $f_{\kappa_a}({t})$ is small, we find the corrections to the so-called orbital and spin $\hat{g}$-factors for the weak external
magnetic field $\bar{B}$. In such a way we demonstrate that the second one of them remains undeformed for any value of
deformation parameter $\kappa_a$.

The paper is organized as follows. In second Section we remaind the basic facts concerning Zeeman effect for hydrogen atom with the electron spin taken into account. In Section 3 we analyze the Zeeman anomaly for quantum space-times (\ref{relations}). Particularly, we find the first-ordered, time-dependent corrections to the spectrum of weak magnetic field as well as, we provide the twist-deformed orbital and spin $\hat{g}$-factors for electron respectively. The discussion and final remarks are presented in the last Section.

\section{The Zeeman effect for hydrogen atom}

In this Chapter we shortly remaind the basic facts concerning Zeeman effect for hydrogen atom with spin electron taken into account, in the case of classical space-time \cite{kiku}. For this purpose, we start with the proper Hamiltonian function\footnote{In the above formulas $m$ and $e$ denote the  mass and charge of the electron respectively, while $c$ is the speed of light.}
\begin{eqnarray}
H(\bar{p},r) = \frac{\bar{p}^2}{2m}+V_{\rm{C}}(r)+V_{\rm{LS}}(r)+H_{\rm{BS}} = H_0+V_{\rm{LS}}(r)+H_{\rm{BS}}\;, \label{ham}
\end{eqnarray}
which contains Coulomb and spin-orbit potentials as well as the spin magnetic-momenta interaction term, given by
\begin{eqnarray}
V_{\rm{C}}(r) &=& -\frac{Ze^2}{{r}}\;, \label{terms1}\\
V_{\rm{LS}}(r) &=& \frac{1}{2m^2c^2}\frac{1}{r}\frac{dV_{\rm{c}}(r)}{dr}\left(\bar{L}\cdot\bar{S}\right)\;, \label{terms2}
\end{eqnarray}
and
\begin{eqnarray}
H_{\rm{BS}} \;=\; -\frac{e}{mc}\bar{B}\cdot\bar{S}\;, \label{terms3}
\end{eqnarray}
respectively. One can check that after the minimal coupling substitution\footnote{$\bar{A}=\frac{1}{2}\left(\bar{B}\times\bar{r}\right)$.}
\begin{eqnarray}
\bar{p} \;\to \; \bar{p}' = \bar{p}-\frac{e}{c}\bar{A}\;, \label{mc}
\end{eqnarray}
the above energy operator takes the form\footnote{${V_{\rm{C}}'(r)}=\frac{d{V_{\rm{C}}(r)}}{dr}$.}
\begin{eqnarray}
H &=& \frac{\bar{p}^2}{2m}-\frac{Ze^2}{{r}}-\frac{e}{2mc}\left(\bar{p}\cdot\bar{A}+\bar{A}\cdot\bar{p}\right)+\frac{e^2\bar{A}^2}{2mc^2}-\frac{e}{mc}
\bar{B}\cdot\bar{S}\;+\;\label{aftermc}\\
&+&{(2m^2c^2)}^{-1}{r}^{-1}{V_{\rm{C}}'(r)}\left(\bar{L}\cdot\bar{S}\right)\;,\nonumber
\end{eqnarray}
while for the magnetic induction chosen into the positive $x_3$-direction $(\bar{B}=B\hat{e}_3)$, it looks as follows
\begin{eqnarray}
H &=& \frac{\bar{p}^2}{2m}+\frac{e^2B^2}{8mc^2}\left(x_1^2+x_2^2\right)-
\frac{e}{2mc}BL_3-\frac{e}{mc}BS_3-\frac{Ze^2}{{r}}\;+ \label{zdirection}\\
&+&{(2m^2c^2)}^{-1}{r}^{-1}{V_{\rm{C}}'(r)}\left(\bar{L}\cdot\bar{S}\right)\;.\nonumber
\end{eqnarray}
Hence, we see, that the total Hamiltonian of the system (\ref{zdirection}) is made up of the four dynamical terms, such that
\begin{eqnarray}
H_{\rm{C}}(r) &=& \frac{{p_3}^2}{2m}-\frac{Ze^2}{{r}}\;, \label{terms1a}\\
H_{\rm{O}}(r) &=& \frac{1}{2m}\left(p_1^2+p_2^2\right)+\frac{e^2B}{8mc^2}\left(x_1^2+x_2^2\right)\;, \label{terms2a}\\
H_{\rm{B}} &=&-\frac{e}{2mc}B\left(L_3+2S_3\right)\;, \label{terms3a}
\end{eqnarray}
and
\begin{eqnarray}
H_{\rm{LS}}(r) ={(2m^2c^2)}^{-1}{r}^{-1}{V_{\rm{C}}'(r)}\left(\bar{L}\cdot\bar{S}\right)=
{(2m^2c^2)}^{-1}\frac{Ze^2}{r^3}\left(\bar{L}\cdot\bar{S}\right)
\;. \label{terms4a}
\end{eqnarray}
First of them describes the electron in the presence of Coulomb field, while the second one corresponds to the electron moving perpendicularly in the standard, two-dimensional harmonic oscillator potential. Besides, the formulas (\ref{terms3a}) and (\ref{terms4a}) encode the Hamiltonian for the anomalous Zeeman effect and spin-orbit interaction terms respectively.

It is well-known, that the first-order corrections to the energy spectrum of the week-field Zeeman anomaly can be found with use of the stationary quantum mechanical perturbation theory \cite{kiku}; they are given by
\begin{eqnarray}
\Delta E_{\rm{weak}} &=& \left<n,l,j,m_j\right|H_{\rm{B}}\left|n,l,j,m_j\right> \;=\;\nonumber\\
&=&-\frac{e\hbar B}{2mc}m_j\left[1+\frac{j(j+1)+s(s+1)-l(l+1)}{2j(j+1)}\right]\;=\;-\frac{e\hbar B}{2mc}m_jg\;, \label{anoamly}
\end{eqnarray}
in an appropriate basis set $\left|n,l,j,m_j\right>$ defined by the principal $(n)$, orbital $(l)$, spin $(s)$ and total $(j=l\pm\frac{1}{2})$
quantum numbers. Besides, it should be noted, that the present in the above formula symbol $g$ denotes so-called Land\'{e} factor, which for $j=l$ and $s=0$ produces the orbital $g$-factor $g_l=1$, while for $j=s$ and $l=0$, it gives the spin Land\'{e} factor $g_s=2$.

\section{The twist-deformed Zeeman effect for hydrogen atom}

Let us now turn to the main aim of our investigations, i.e., to the analyze of Zeeman anomaly for quantum space-times (\ref{relations}).
In first step of our construction, we extend the twisted space to the whole algebra of momentum and position operators as follows
\begin{eqnarray}
&&[\;\hat{ x}_{1},\hat{ x}_{2}\;] = if_{\kappa_a}({t})\;\;\;,\;\;\;
[\;\hat{ p}_{i},\hat{ p}_{j}\;] = 0 = [\;\hat{ x}_{i},\hat{ x}_{3}\;]\;\;\;,\;\;\;[\;\hat{ x}_{i},\hat{ p}_{j}\;] =
i\hbar\delta_{ij}\;.\label{rel1}
\end{eqnarray}
One can check that the above relations satisfy the Jacobi identity and for deformation parameter
$\kappa_a$ approaching zero become classical. \\
Next, we define the proper Hamiltonian operator in a standard way by\footnote{We define the Hamiltonian operator by replacement in the formula (\ref{zdirection}) the classical operators $(x_i,p_i)$ by their noncommutative counterparts $(\hat{x}_i,\hat{p}_i)$.}
\begin{eqnarray}
\hat{H} &=& \frac{\bar{\hat{p}}^2}{2m}+\frac{e^2B^2}{8mc^2}\left(\hat{x}_1^2+\hat{x}_2^2\right)-
\frac{e}{2mc}BL_3-\frac{e}{mc}BS_3-\frac{Ze^2}{{\hat{r}}}\;+ \label{twistzdirection}\\
&+&{(2m^2c^2)}^{-1}{V'_{\rm{C}}(\hat{r})}{\hat{r}}^{-1}\left(\bar{L}\cdot\bar{S}\right)\;,\nonumber
\end{eqnarray}
and in order to perform the basic analyze of the system, we represent the
noncommutative variables $({\hat x}_i, {\hat p}_i)$ by the classical
ones $({ x}_i, { p}_i)$ as  (see e.g.
\cite{romero1}-\cite{kijanka})
\begin{eqnarray}
{\hat x}_{1} &=& { x}_{1} - {{f_{\kappa_a}(t)}}/{(4\hbar)}p_2\;\label{rep1}\\
{\hat x}_{2} &=& { x}_{2} + {{f_{\kappa_a}(t)}}/{(4\hbar)}p_1
\;,\label{rep2}\\
{\hat x}_{3} &=& { x}_{3}\;\;\;,\;\;\;{\hat p}_{i} \;=\; { p}_{i} \;,
\label{rep3}
\end{eqnarray}
where
\begin{equation}
[\;x_i,x_j\;] = 0 =[\;p_i,p_j\;]\;\;\;,\;\;\; [\;x_i,p_j\;]
={i\hbar}\delta_{ij}\;. \label{classpoisson}
\end{equation}
Then, the  Hamiltonian (\ref{twistzdirection}) takes the form
\begin{eqnarray}
\hat{H}(t) &=&
\frac{1}{2\hat{m}(t)}\left({{{p}}_1^2}+{{{p}}_2^2} \right)  + \frac{{{p}}_3^2}{2m}+
{{\hat{m}(t)\hat{\omega}^2(t)}}/{2}\left({{{x}}_1^2}+{{{x}}_2^2} \right)-\left[
S(t)+\frac{eB}{2mc}\right]L_3\;+ \label{2dh1}\\
&-&\frac{e}{mc}BS_3-\frac{Ze^2}{{\hat{r}(x,p)}}+
{(2m^2c^2)}^{-1}{V_{\rm{C}}'(\hat{r}(x,p))}{\hat{r}(x,p)}^{-1}\left(\bar{L}\cdot\bar{S}\right)\;,\nonumber
\end{eqnarray}
with time-dependent mass and frequency given by
\begin{eqnarray}
\hat{m}(t)&=&\frac{m}{1+{e^2B^2f_{\kappa_a}^2(t)}/{(64c^2\hbar^2)}}\;,\\
\hat{\omega}(t)&=&
\frac{eB}{2{m}c}\left[1+\frac{e^2B^2f_{\kappa_a}^2(t)}{64c^2\hbar^2}\right]^{\frac{1}{2}}\;,\\
S(t)&=&\frac{\hat{m}(t)\hat{\omega}^2(t)f_{\kappa_a}(t)}{4\hbar}L_3\;,
\end{eqnarray}
respectively. Besides, by tedious calculations one can check, that for small values of deformation function $f_{\kappa_a}(t)$, the total energy operator (\ref{2dh1}) is made up from the following four terms
\begin{eqnarray}
\hat{H}_{\rm{C}}(r) &=& \frac{{p_3}^2}{2m}-\frac{Ze^2}{{r}}\;, \label{terms1aa}\\
\hat{H}_{\rm{O}}(r) &=& \frac{1}{2m}\left(p_1^2+p_2^2\right)+\frac{e^2B^2}{8mc^2}\left(x_1^2+x_2^2\right)+{\cal O}(f_{\kappa_a}^2(t))\;, \label{terms2aa}\\
\hat{H}_{\rm{B}}(t) &=&A (t)L_3-{eB}/{(mc)}S_3+{\cal O}(f_{\kappa_a}^2(t))\;, \label{terms3aa}\\
\hat{H}_{\rm{LS}}(r,t) &=&({2m^2c^2})^{-1}({Ze^2}{r^{-3}})\left[1+{(3f_{\kappa_a}(t)}/{4\hbar}){L_3}{r^{-2}}\right]\left(\bar{L}\cdot\bar{S}\right)\;+
\; \label{terms4aa}\\
&+&{\cal O}(f_{\kappa_a}^2(t))\;,
\end{eqnarray}
where symbol ${\cal O}(f_{\kappa_a}^2(t))$ denotes the elements at least quadratic in parameter $\kappa_a$, while the time-dependent coefficient $A(t)$ is given by
\begin{eqnarray}
A(t)=-S(t)-\frac{Zf_{\kappa_a}(t)}{4\hbar r^3}-\frac{eB}{2mc}\;.
\end{eqnarray}
Of course, in $f_{\kappa_a}(t)$ approaching zero limit the above formulas become the same as classical ones (\ref{terms1a})-(\ref{terms4a}).

In the next step of our construction one should observe, that formally, the first-order Zeeman anomaly corrections to hydrogen spectrum in a week external magnetic field can be written as\footnote{In our calculation of formula (\ref{anoamlync}) we consider the following perturbation of eigenvalues $E_n(t)$ and eigenvectors $\left|n,l,j,m_j\right>(t)$ respectively $$E_n(t)=E_n^{(0)}+\lambda E_n^{(1)}(t)+\cdots\;\;\;,\;\;\;\left|n,l,j,m_j\right>(t)=\left|n,l,j,m_j\right>+\lambda \left|n,l,j,m_j\right>^{(1)}(t)+\cdots\;,$$ where $H_0\left|n,l,j,m_j\right>=E_n^{(0)}\left|n,l,j,m_j\right>$.}
\begin{eqnarray}
E_n^{(1)}(t)=\Delta \hat{E}_{\rm{weak}}(t) = \left<n,l,j,m_j\right|\hat{H}_{\rm{B}}(t)\left|n,l,j,m_j\right>  \;, \label{anoamlync}
\end{eqnarray}
and due to the relations \cite{kiku}
\begin{eqnarray}
\left<L_3\right>&=&
\hbar m_j\left[\frac{j(j+1)-s(s+1)+l(l+1)}{2j(j+1)}\right]\;, \label{exp1}\\
\left<S_3\right>&=&
\hbar m_j\left[\frac{j(j+1)+s(s+1)-l(l+1)}{2j(j+1)}\right]\;, \label{exp2}
\end{eqnarray}
they take the form
\begin{eqnarray}
\Delta \hat{E}_{\rm{weak}}(t)&=&-\frac{eB\hbar}{2mc} m_j\left[\left[\frac{2mc}{eB}\left(S(t)+\frac{Zef_{\kappa_a}(t)}{4\hbar}\left<\frac{1}{r^3}\right>\right)+1\right. \right]\times \nonumber\\
&\times&\left(\frac{j(j+1)-s(s+1)+l(l+1)}{2j(j+1)}\right)\;+\label{anoamlync1x}\\
&+&2\left.\left(\frac{j(j+1)+s(s+1)-l(l+1)}{2j(j+1)}\right)\right]\;=\;-\frac{eB\hbar}{2mc}m_j\hat{g}(t)\;,
\end{eqnarray}
with $\left<\frac{1}{r^3}\right>=\frac{Z^3}{l(l+1/2)(l+1)n^3a_0^3}$ being a Kramer's recursive relation, $a_0=\frac{\hbar^2}{me^2}$, and with $\hat{g}(t)$ denoting the Land\'{e} factor for twist-deformed space-time. Unfortunately, the first term in the above formula remains indeterminate due to the singularity of expectation value $\left<{r^{-3}}\right>$ for $l=0$. In order solve this problem one should exchange (in accordance with articles \cite{36} and \cite{qwerty}) the Kramer's factor to the following one: $\left<{r^{-3}}\left(1-{a_0\alpha^2}/{2r}\right)\right>$ with $\alpha=e^2/(\hbar c)$. Then, the pure orbital moment Land\'{e} factor can be get from (\ref{anoamlync1x}) by taking $j=l$ and $s=0$, and it looks as follows\footnote{Particularly, for $f_{\kappa_a}(t)=\kappa_a=\theta$ we obtain the result for canonically deformed space-time \cite{oeckl}-\cite{dasz1}.}
\begin{eqnarray}
\hat{g}_l(t)=\frac{2mc}{eB}\left(S(t)+\frac{Zef_{\kappa_a}(t)}{4\hbar}\left<r^{-3}\left(1-{a_0\alpha^2}/{2r}\right)\right>\right)+1\;, \label{pureor}
\end{eqnarray}
while for $f_{\kappa_a}(t)=0$, it reproduces the commutative ${g}$-factor $g_l=1$. Besides, by putting $j=s$ and $l=0$ in (\ref{anoamlync1x}) we obtain the pure spin moment of the form
\begin{eqnarray}
\hat{g}_s(t)=2=g_s \;, \label{purespin}
\end{eqnarray}
what finish our considerations.

\section{Final remarks}

As it was already mentioned in Introduction, in this article, we investigate the impact of twisted  space-time  (\ref{relations})
on the energy spectrum of hydrogen atom in a weak magnetic field, with the electron spin taken into account. Particularly, we find the linear (with respect the noncommutativity function $f_{\kappa_a}({t})$) corrections to the orbital and spin Zeeman $\hat{g}$-factors (\ref{pureor}) and (\ref{purespin}) respectively. Surprisingly, we notice that the second one of them remains undeformed.

Of course, the presented and discussed in this article results can be generalized to the case of arbitrary (for example much more strong) magnetic field $\bar{B}$ as well. Unfortunately, the studies in this direction seem to be very complicated and for this reason they are postponed for further investigations.

\section*{Acknowledgments}

The author would like to thank J. Lukierski for valuable discussions.

\end{document}